\documentclass[fleqn,usenatbib]{mnras}

\usepackage{newtxtext,newtxmath}

\usepackage[T1]{fontenc}

\DeclareRobustCommand{\VAN}[3]{#2}
\let\VANthebibliography\thebibliography
\def\thebibliography{\DeclareRobustCommand{\VAN}[3]{##3}\VANthebibliography}

\usepackage{graphicx}	
\usepackage{amsmath}	
\usepackage[normalem]{ulem}

\newcommand{\capcolor}[1]{{\color{blue}#1}}
\graphicspath{{./}{figures/}}

\title[20 minute time lag observed from Sgr A* at X band]{Detection of a 20 minute time lag observed from Sgr A* between 8 and 10 GHz with the VLA}

\author[J. M. Michail, F. Yusef-Zadeh, \& M. Wardle (2021)]{
Joseph M. Michail,$^{1}$\thanks{E-mail: michail@u.northwestern.edu}
Farhad Yusef-Zadeh,$^{1}$
and Mark Wardle$^{2}$
\\
$^{1}$Center for Interdisciplinary Exploration and Research in Astrophysics (CIERA) and Department of Physics and Astronomy, Northwestern University,\\ 1800 Sherman Avenue, Evanston, IL. 60201, USA\\
$^{2}$Research Centre for Astronomy, Astrophysics and Astrophotonics and Department of Physics and Astronomy, Macquarie University, Sydney, NSW 2109, Australia
}

\date{Accepted 2021 May 24. Received 2021 May 4; in original form 2021 March 21}

\pubyear{2021}

\begin{document}
\label{firstpage}
\pagerange{\pageref{firstpage}--\pageref{lastpage}}
\maketitle

\begin{abstract}
    We report the detection and analysis of a radio flare observed on 17 April 2014 from Sgr A* at $9$ GHz using the VLA in its A-array configuration. This is the first reported simultaneous radio observation of Sgr A* across $16$ frequency windows between $8$ and $10$ GHz. We cross correlate the lowest and highest spectral windows centered at $8.0$ and $9.9$ GHz, respectively, and find the $8.0$ GHz light curve lagging $18.37^{+2.17}_{-2.18}$ minutes behind the $9.9$ GHz light curve. This is the first time lag found in Sgr A*'s light curve across a narrow radio frequency bandwidth. We separate the quiescent and flaring components of Sgr A* via flux offsets at each spectral window.
    The emission is consistent with an adiabatically-expanding synchrotron plasma, which we fit to the light curves to characterize the two components. The flaring emission has an equipartition magnetic field strength of $2.2$ Gauss, size of $14$ Schwarzschild radii, average speed of $12000$ km s$^{-1}$, and electron energy spectrum index ($N(E)\propto E^{-p}$), $p = 0.18$. The peak flare flux at $10$ GHz is approximately $25$\% of the quiescent emission. 
    This flare is abnormal as the inferred magnetic field strength and size are typically about $10$ Gauss and few Schwarzschild radii. The properties of this flare are consistent with a transient warm spot in the accretion flow at a distance of $10$-$100$ Schwarzschild radii from Sgr A*.
    Our analysis allows for independent characterization of the variable and quiescent components, which is significant for studying temporal variations in these components. 
\end{abstract}




\section{Introduction}\label{sec:intro}

    Sagittarius A* (Sgr A*) is the closest example of a supermassive black hole, located at the dynamical center of the Galaxy at a distance of approximately $8.2$ kpc \citep{Gravity2019}. Sgr A* was discovered by \cite{Balick1974} with the Green Bank Interferometer and subsequently confirmed by the Westerbork Array and Very Long Baseline Interferometry \citep[see][and references therein]{Melia2001}, \cite{Brown1982} detected its radio variability shortly thereafter. 
    
    The proximity of this source to the Earth has made it a prime target for a close-up view of how supermassive black holes at the center of other galaxies interact with their surroundings. As such, Sgr A* has been well monitored at wavelengths from the radio to X-rays, all of which show flux variability. At infrared wavelengths, there are about four daily flares \citep[and references therein]{DoddsEden2009}, which are likely optically-thin synchrotron emission \citep{Eckart2004, Eckart2006}. At radio and submillimeter wavelengths, flaring occurs on hourly timescales \citep{Zadeh2011, Dexter2014, Iwata2020}. Lifetimes of such flares are shorter than radio/submillimeter synchrotron cooling times, and are consistent with an adiabatically-expanding synchrotron plasma \citep{Zadeh2006b}. The variable emission present at all wavelengths is considered part of the "flaring component," the origin of which is unknown and might originate from outflows, jets, or accretion flows.
    
    Sgr A* also has a steady or quasi-steady component, known as the "quiescent component," which has been studied in the radio and submillimeter. The emission from this component is likely dominated by an accretion disk. \cite{Herrnstein2004} noted epochs of varying flux and spectral index in the radio, which might indicate different states of accretion onto the black hole. Studying the quiescent component in the radio and submillimeter is difficult, as the variability of the flaring component superimposes itself onto the more slowly varying quiescent emission. Therefore, radio observations of Sgr A* intrinsically measure the sum of the fluxes from these components.

    In an attempt to study only the quiescent component, time-averaging the emission in this wavelength regime has been used. \citet{Duschl1994} use radio and submillimeter observations averaged over $11$ years to determine the spectrum of Sgr A*. Their analysis yields a power-law of the form $S_\nu \propto \nu^{0.35-0.38}$. Similar analyses find different spectral indices consistent with the range $S_\nu \propto \nu^{0.1 - 0.4}$ \citep[][and references therein]{Melia2001}. However, comparable to the result of \cite{Herrnstein2004} noted above, additional long-term variability studies have been completed, which shows that this technique may not be appropriate. \cite{Falcke1999} identifies variability at $2.3$ and $8.3$ GHz on timescales between $50$ and $200$ days. Additional analyses at $23$ GHz suggests a $106$ day periodicity \citep{Zhao2001}. However, a reanalysis of this data by \citet{Macquart2006} was unable to confirm this value. Therefore, to study the quiescent component, the flaring component must be first characterized and removed.

    Simultaneous multi-wavelength observations at radio, submillimeter, infrared, and X-ray wavelengths have been used to characterize the flaring component \citep[e.g,][]{Falcke1998, An2005, Zadeh2008, Zadeh2009, Mossoux2016}. Radio observations using this technique are particularly numerous from the historic VLA, with fast-frequency switching enabling near-simultaneous observations at $22$ and $43$ GHz \citep[e.g.][]{Zadeh2006, Zadeh2008}. \cite{Zadeh2006} used an adiabatically-expanding synchrotron plasma model \citep["plasmon" model,][]{VDL1966} to characterize a flare from Sgr A*, where the peak flux at $22$ GHz was delayed by approximately $20$ minutes relative to $43$ GHz. This model naturally describes time delays between different pairs of frequencies as decreasing optical depth of the plasma blob with increasing size and has been subsequently applied to many flaring events observed across the electromagnetic spectrum \citep[e.g.,][]{Eckart2006,  Eckart2008, Marrone2008, Zadeh2008, Zadeh2009, Miyazaki2013}. 
    
    The upgraded VLA correlator now allows for truly simultaneous observations across several GHz of bandwidth in the radio spectrum, providing opportunities to apply this model without fast-frequency switching or observations from other telescopes. Therefore, a single VLA frequency band with the same UV coverage can be used to calculate properties of the flaring plasma, such as electron energy index, magnetic field strength, electron density, and expansion speed.
    
    Here, we present an analysis of a radio flare from Sgr A* observed with the Karl G. Jansky Very Large Array (VLA) on 17 April 2014. We report a strong flare lasting at least $6$ hours, which was observed at central frequency of $9$ GHz with $2$ GHz of total bandwidth.  In Section \ref{sec:reduction}, we present our data reduction methods. In Section \ref{sec:results}, we present cross correlation analyses between the light curves of the most widely separated spectral windows. We attempt to split the flaring and quiescent components via constant offsets at each of the spectral windows. In Section \ref{sec:discussion}, we characterize the quiescent and flaring components to this remarkable observation by simultaneously fitting a sum of adiabatic expansion and power-law models across each of the $16$ spectral windows for the first time. We subsequently compare the results from the constant offset model, those from this more detailed analysis, and from previous work. In Section \ref{sec:summary}, we summarize our results.

\section{Observation and Data Reduction}\label{sec:reduction}

    The VLA observed Sgr A* on 17 April 2014 as part of project ID 14A-232 (PI: Yusef-Zadeh) in the A-configuration for a total on-source time of approximately $5.5$ hours. Observations completed on this date used the X-band receivers centered at $9$ GHz with a total bandwidth of approximately $2$ GHz. The total bandwidth was composed of $16$ intermediate frequencies (IFs), each with a bandwidth of $128$ MHz. Additionally, each IF was composed of $64$ channels with a bandwidth of $2$ MHz. This frequency configuration allowed for observing the $4$ circular polarization products (RR, LL, RL, and LR). The reduction of this data set is described below.
        
    The observation is composed of subsequent scans of calibrator targets (3C286, J1733-1304, and J1744-3116) and the science target, Sgr A*. 3C286, a standard VLA calibrator, is used to set the flux scaling and polarization angle. J1733-1304 is used as the bandpass calibrator. J1744-3116 (denoted J1744 throughout the rest of this paper) is used as the complex gain and polarization leakage calibrator.
        
    Data for this day's observation were downloaded from the VLA archive in the native SDM-BDF format. We reduced this data set with the default VLA pipeline within \texttt{CASA} \citep[version 5.6.2]{McMullin2007}. We follow the default \texttt{CASA} polarization reduction\footnote{\url{https://casaguides.nrao.edu/index.php?title=CASA_Guides:Polarization_Calibration_based_on_CASA_pipeline_standard_reduction:_The_radio_galaxy_3C75-CASA5.6.2}} to properly calibrate the polarimetry. The Sgr A* and J1744 data were exported to UVFITS files and imported into \texttt{AIPS} for further processing. \texttt{TVFLG} was used to flag additional misbehaving baselines and antennas not flagged in the initial rounds of processing. Phase self-calibration was completed on both targets to remove effects of atmospheric turbulence. These self-calibrated data are used with \texttt{AIPS} task \texttt{DFTPL} to make light curves of both Sgr A* and J1744, where baselines greater than $100 \text{ k}\lambda$ are used with a $1$ minute binning time for each IF. Discussion of the polarization analysis is left for a future paper.

     Throughout our analysis, we noticed flux discontinuities present between neighboring IFs, which cause the spectrum of Sgr A* to deviate from a pure power law. The jumps in Sgr A*'s spectrum are identical to the flux bootstrapping residuals for J1733-1304 and J1744-3116 in the CASA pipeline. The residual fluxes of the two calibrators are on the order of $1\%$, which match the expected accuracy of this pipeline step\footnote{\url{https://science.nrao.edu/facilities/vla/docs/manuals/oss2014A/performance/fdscale}}. These discontinuities are notably present for IFs toward the ends of a baseband since they are plagued by decreased sensitivity from spectral roll-off.
     
\section{Results}\label{sec:results}

        \begin{figure}
            \centering
             \includegraphics[trim = 1.0cm 1.55cm 1.0cm 1.8cm, clip,width=\columnwidth]{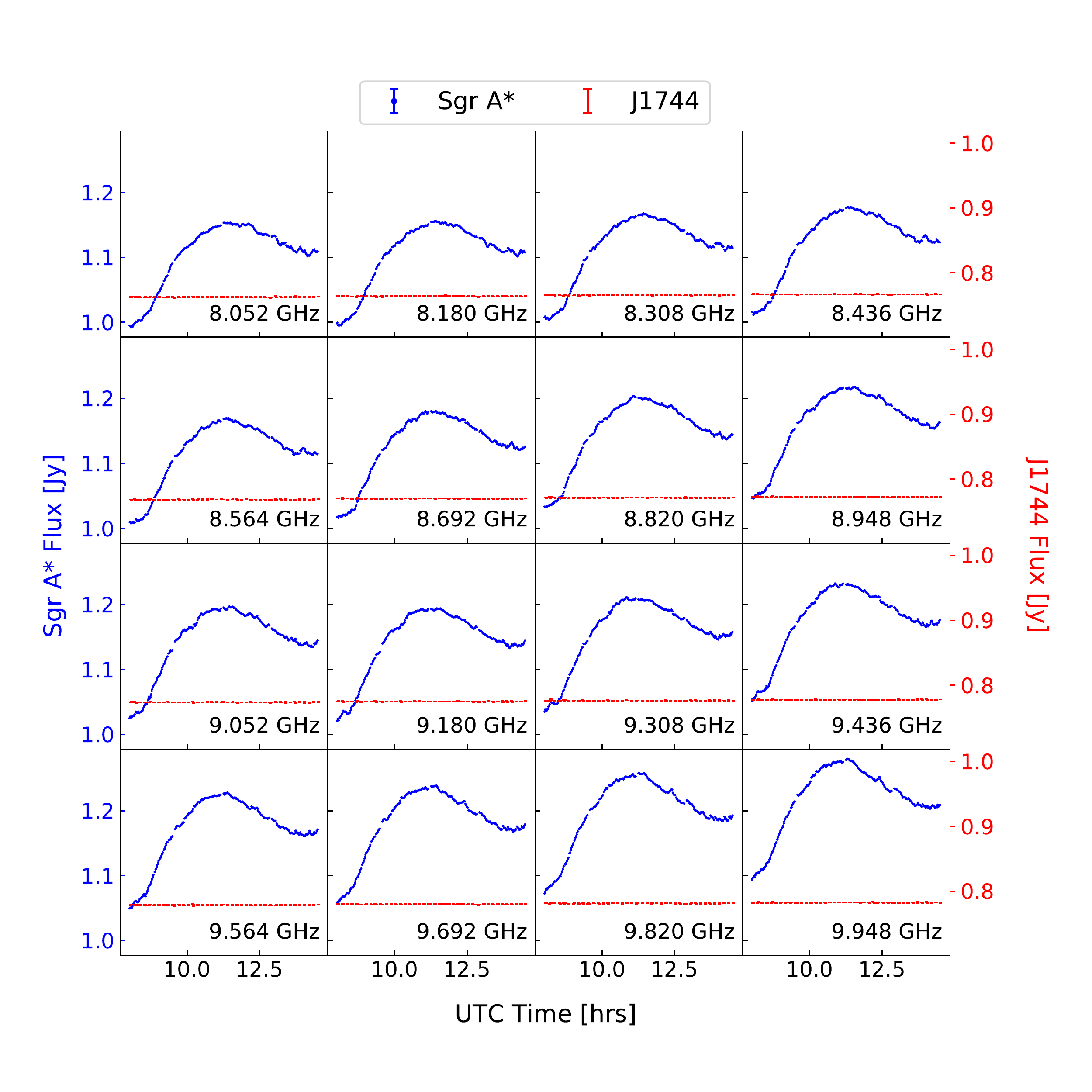}
            \caption{Total intensity light curves on 17 April 2014 for Sgr A* (in blue) and J1744-3116 (red) for the $16$ intermediate frequencies (IFs) discussed in this paper. Note that the light curves for the two sources are on different scales. Error bars for the Sgr A* data are smaller than the markers, whereas error bars for J1744 are about the size of the markers.}
            \label{fig:flux}
        \end{figure}
        
        In Figure \ref{fig:flux} we present the $16$ simultaneous light curves of Sgr A* (in blue) and J1744 (in red). The frequency of each IF is shown at the bottom right of each panel. All of the light curves show the same general trend of increasing flux at a level of approximately $15\%$ from the beginning of the observation with a slow decay after the peak occurs. This cannot be an instrumental effect as the light curve of J1744 remains constant throughout the observation. 
        
        \subsection{Cross Correlation Analysis}\label{ssec:timedelay}
\begin{figure*}
                \centering
                \includegraphics[width=\textwidth]{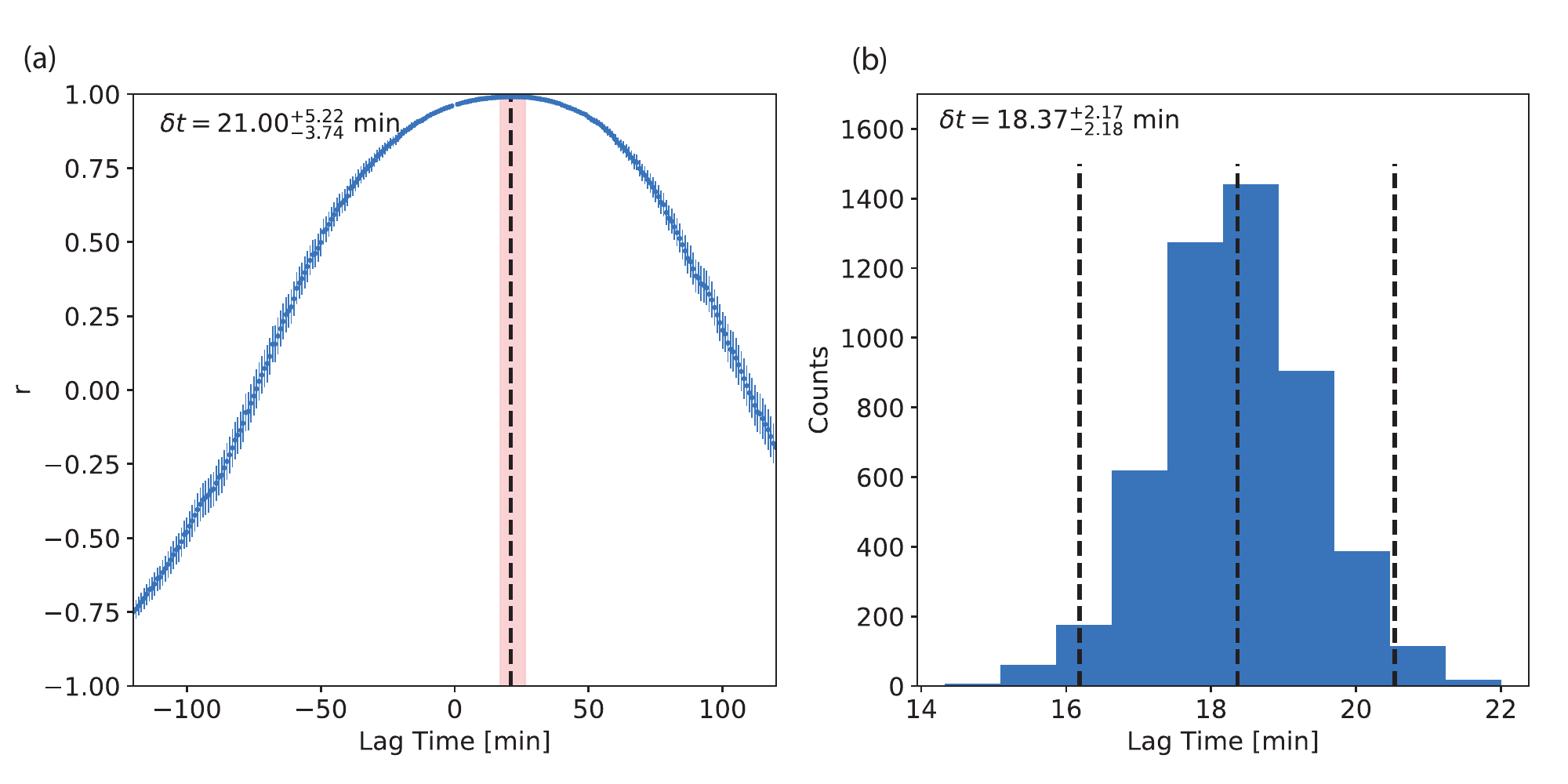}
                \caption{({\color{blue} a}) The cross correlation function using the z-transformed discrete correlation function \citep[ZDCF; ][]{Alexander1997} between $8.052$ and $9.948$ GHz. The analysis finds that the lowest frequency data peaks about $20$ minutes after the highest frequency data. A $2\sigma$ confidence interval, shaded in pink, is used for the errors. ({\color{blue} b}) The histogram of centroid lag times using $5000$ Monte Carlo simulations from the \texttt{pyCCF} \citep{PYCCF} Python package between the highest and lowest frequency light curves. This method finds the highest frequency light curve peaking about $18$ minutes ahead of the lowest frequency, comparable to the time lag found using the ZDCF method. A $2\sigma$ error range is shown.}
                \label{fig:cross_corr}
            \end{figure*}

            We begin by completing a cross correlation analysis between the most widely separated frequencies to determine if there is a time delay between light curves across the band. This is performed in two ways. The first uses the z-transformed discrete correlation function \citep[ZDCF; ][]{Alexander1997}, which is especially useful for un-evenly sampled light curves. In conjunction with the \texttt{PLIKE} program \citep{Alexander2013}, an error interval on the peak time-lag can be determined. For this analysis, we use a $95\% (2\sigma)$ confidence interval. We show the cross correlation function (CCF) between the $8.052$ and $9.948$ GHz light curves in Figure \ref{fig:cross_corr}\capcolor{a}, where we find the lowest frequency to peak $21.00^{+5.22}_{-3.74}$ minutes after the highest frequency. Note that this value is comparable to the typical time delay observed between $22$ and $43$ GHz of about $20$ minutes \citep[][]{Zadeh2006, Zadeh2008}.
            
\begin{figure}
                \centering
                \includegraphics[trim= 0 2.0cm 0 2.2cm, clip, width=\columnwidth]{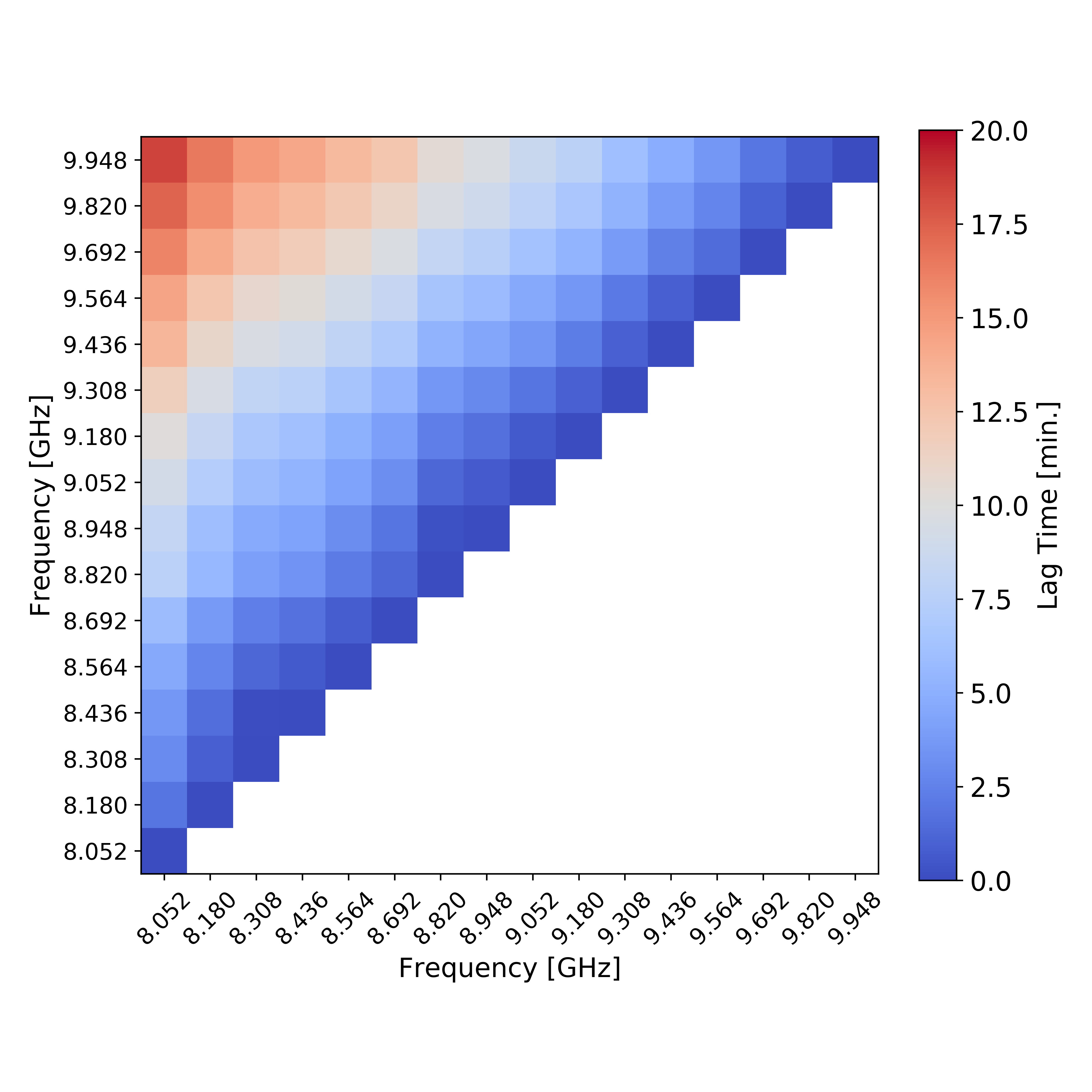}
                \caption{A plot of all measured time delays between each pair of frequencies. The light curve for any frequency on the x-axis peaks after the light curve on the y-axis. The color displays the time delay between these two frequencies as determined by the \texttt{pyCCF} package.}
                \label{fig:waterfall}
            \end{figure}

            As an additional check, we use the Python package \texttt{pyCCF} \citep{Peterson1998, PYCCF} for an independent measure of the time delay. This package determines the time lag between two datasets via the interpolated CCF function, where two light curves at irregularly-sampled times are interpolated to the same time grid such that typical cross correlation analyses can be completed. Monte Carlo simulations are used to build a histogram of time lags to find the mean lag and error interval. The cross correlation centroid peak distribution is shown in Figure \ref{fig:cross_corr}\capcolor{b} with a $2\sigma$ interval denoted. Again, we find that the lower frequency peaks roughly $20$ minutes after the highest frequency, confirming that this is not a spurious result.
            Due to the wide simultaneous frequency coverage provided by the VLA, we determine the time lag for each pair of spectral windows. This is shown in Figure \ref{fig:waterfall}, where the \texttt{pyCCF} package was used to calculate the lag. For any pair of frequencies, the figure shows that the light curve at a frequency on the x-axis peaks after the light curve at a higher frequency on the y-axis. This can be confirmed by considering the measured $18.37_{-2.18}^{+2.17}$ minute delay between $8.052$ and $9.948$ GHz. The smooth gradient of measured lags between the most- and least-widely separated frequencies demonstrates that the time lag presented above is not due to spurious features in the light curves. Instead, it is an intrinsic property of them.
        \subsection{Spectral Index of the Quiescent and Flaring Components}\label{ssec:brute}
            The observed flux at any frequency, denoted $S_\nu$, from Sgr A* is a superposition of the quiescent and flaring components. \cite{Zadeh2006} attempted to separate these components by removing a steady flux offset from both observed frequencies. This zeroth-order method allows for general properties of the quiescent and flaring components (such as spectral index) to be calculated. Here, the flux offset for each frequency is determined at the beginning of the observation. This assumes the flare observed starts in concert with the observations, which, while unlikely, allows us to study a heuristic model of this observation. We denote this the "constant offset model" throughout this paper.
            
            We note that an inaccurate decomposition of the two components will affect the results of this model; however, they strongly affect the flaring component. This is caused by the relative fluxes of the two components, as the quiescent emission is much stronger than the variable emission. 
            
\begin{figure*}
                \centering
                \includegraphics[width=\textwidth]{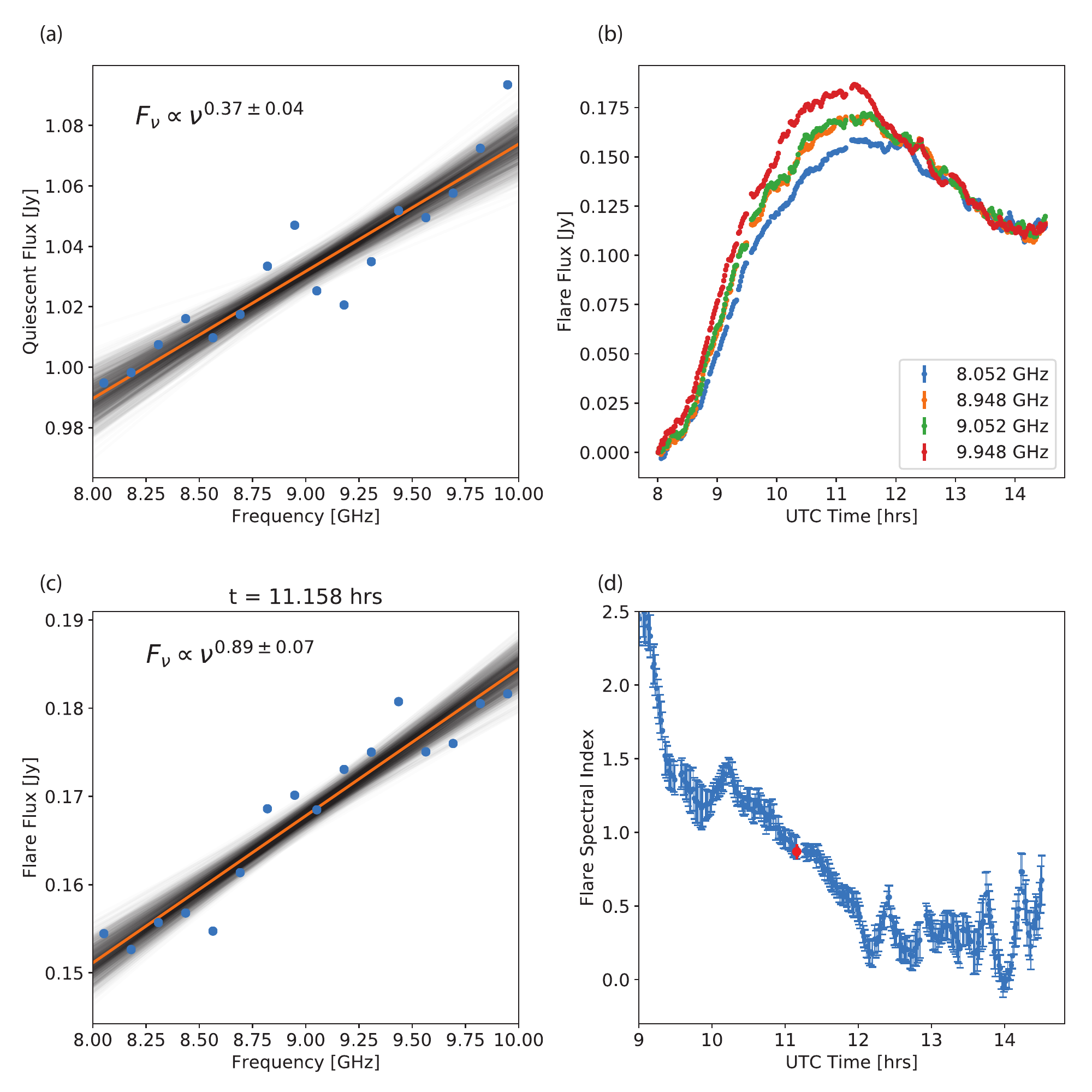}
                \caption{({\color{blue} a}) The spectral energy distribution (SED) of the quiescent component assuming a constant flux in time. The orange line denotes the spectral index fit of the data using an error-weighted linear-least squares regression. The black lines are $1000$ Monte Carlo realizations of the fit. ({\color{blue} b}) The light curves of the flaring component for four of our observed frequencies. ({\color{blue} c}) The SED of the flare at the time of peak flare emission at $9.948$ GHz. The orange line denotes the spectral index fit of the flare at this instant using an error-weighted linear-least squares regression. The black lines are $1000$ Monte Carlo realizations of the fitted spectral index. ({\color{blue} d}) Spectral index in time for the flaring component of Sgr A*. The red point denotes the time of peak flux at $9.948$ GHz.}
                \label{fig:quiescent_params}
            \end{figure*}

            Following \citet{Zadeh2006}, we subtract the flux from each of the IFs at the beginning of the observation and denote these values the quiescent fluxes. They range from $0.99$ Jy at $8.052$ GHz to $1.09$ Jy at $9.948$ GHz. The residual fluxes are the flare light curves. The spectral index is defined as:
                \begin{equation}
                    \centering
                    \alpha \equiv \dfrac{d\log S_\nu}{d\log\nu}.
                \end{equation}
            An error-weighted linear-least squares regression is used to fit the spectral index. In Figure \ref{fig:quiescent_params}\capcolor{a}, we show the spectral energy distribution (SED) of the quiescent component with the fitted spectral index. The light curves of the flare at four representative frequencies are shown in Figure \ref{fig:quiescent_params}\capcolor{b}.
            
            There is structure present in the quiescent SED (Figure \ref{fig:quiescent_params}\capcolor{a}) centered around $9.0$ GHz, which deviates from a pure power-law at a level greater than $1\sigma$. This is caused by imperfect flux bootstrapping from the calibrators to Sgr A* (see Section \ref{sec:reduction} for additional details).
            
            We determine the spectral index of the flaring component over time using the flare-only light curves, which are shown in Figure \ref{fig:quiescent_params}\capcolor{d} for $t > 9$ hours UTC. Times before $9$ hours UTC are highly variable and are strongly dependent upon the choice of quiescent flux chosen in the original light curves; therefore, we do not use these points in this analysis. Toward the beginning of the flare, the spectral index is approximately $\alpha \approx 2.5$, indicative of optically-thick synchrotron emission. At the end of the observation, the flare spectral index levels out near $\alpha \approx 0.3$. The red point shown in Figure \ref{fig:quiescent_params}\capcolor{d} marks the time at which the $9.948$ GHz light curve reaches its maximum flux. An example of the flare SED at that time is shown in Figure \ref{fig:quiescent_params}\capcolor{c}.

            Our analysis finds the spectral index of the quiescent component is $\alpha_c = 0.37 \pm 0.04$ while the spectral index of the flare emission at the $9.948$ GHz peak is approximately twice as steep at $\alpha_f = 0.89 \pm 0.07$. The flare characteristics here are comparable to the flare studied in \cite{Zadeh2006}, where the authors find the peak flare spectral index is roughly double the quiescent component spectral index. 
            
            The temporal variations of the flaring emission's flux and spectral index are consistent with an adiabatically-expanding plasma model as described in \citet{Zadeh2006}, which warrants a complete analysis of this flare using this picture.
            
\section{Discussion}\label{sec:discussion}
        \subsection{The Adiabatic Expansion Model}\label{ssec:plasmon}
            The adiabatic expansion picture begins with a uniform, spherical blob of synchrotron-emitting plasma that is optically-thick at radio wavelengths. It is composed of electrons with a power-law number density of the form $N(E)\propto E^{-p}$ with upper and lower energy bounds $E_1$ and $E_2$, which are functions of time. It expands adiabatically and is assumed to not interact with a surrounding medium, neither transferring magnetic flux nor energy density. The expansion of the blob decreases the optical depth until it becomes optically-thin, which causes the light curves to reach a peak flux then decay. This occurs at progressively lower frequencies as the blob expands, which naturally explains the time delays between the peak fluxes at different frequencies. 
            
            The model is fully-characterized by five parameters: $S_0^p, \beta, t_0$, $r$, and $p$. $S_0^p$ is the peak flux at reference frequency $\nu_0$. $\beta$ is the acceleration parameter of the blob, where values of $\beta > 1$ indicate acceleration, $\beta < 1$ deceleration, and $\beta = 1$ is uniform expansion. $t_0$ is defined as the "apparent age" of the source by \cite{VDL1966} under the assumption of constant acceleration. $r$ is the normalized blob radius and is related to $\beta$ and $t_0$ via the relation: $r \equiv (t / \beta t_0)^\beta$. $t$ is defined as the age of the flare. $p$ is the power-law index of the electron energy spectrum ($E^{-p}$). The flux density at any frequency and time is described via:
            \begin{equation}
                S_p(\nu, r) = S_0^p \left(\dfrac{\nu}{\nu_0}\right)^{5/2}r^3\left(\dfrac{1-\exp({-\tau_\nu})}{1-\exp({-\tau_0})}\right),
                \label{eq:plasmon}
            \end{equation}
            where $\tau_\nu$ is the optical depth at frequency $\nu$ defined as
            \begin{equation}
                \tau_\nu = \tau_0\left(\dfrac{\nu}{\nu_0}\right)^{-(p+4)/2}r^{-(2p+4)}.
            \end{equation}
            $\tau_0$, the optical depth of the flare at its peak, satisfies the following relation:
            \begin{equation}
                e^{\tau_0} - \left(\dfrac{2p}{3}+1\right)\tau_0 - 1 = 0,
            \end{equation}
        \citep{Zadeh2006}. In Section \ref{ssec:brute}, we assume that the quiescent flux of Sgr A* is constant during the observation. However, a sophisticated model will be able to distinguish between a time-independent or -dependent quiescent flux. To first order, the time-dependent emission of the quiescent flux can be described by:
            \begin{equation}
                S_q(\nu, t) = \left(S_0^q + \dot{S}_q(t-8)\right)\left(\dfrac{\nu}{\nu_0}\right)^{\alpha_q}.
                \label{eq:quiescent}
            \end{equation}
        Here, $t$ is the UTC time, $S_0^q$ is the flux at frequency $\nu_0$ at $t=8$ hours UTC, $\dot{S}_q$ is the frequency-independent time derivative of $S_0^q$, and $\alpha_q$ is the spectral index of the quiescent component. The universal time is taken relative to the start of the observation ($t\approx 8$ hours UTC). In this form, Equation \ref{eq:quiescent} is a first-order Taylor expansion of the quiescent component's flux in time. This more general case places no constraints on the quiescent component's possible variable nature, which will affect the fitted flaring model parameters. The use of a time-variable quiescent component was presented in \cite{Zadeh2008} to simultaneously fit flaring emission from Sgr A* at $22$ and $43$ GHz. Adopting an explicit time-dependent term, whose strength is a free parameter, allows for an unbiased determination of the quiescent component's underlying nature. If the quiescent flux is constant, $\dot{S}_q\approx0$, and the flux at any frequency is only a power-law. This form additionally guarantees a constant spectral index for the quiescent component over the observation.
        
\begin{figure*}
                \centering
                \includegraphics[width=\textwidth]{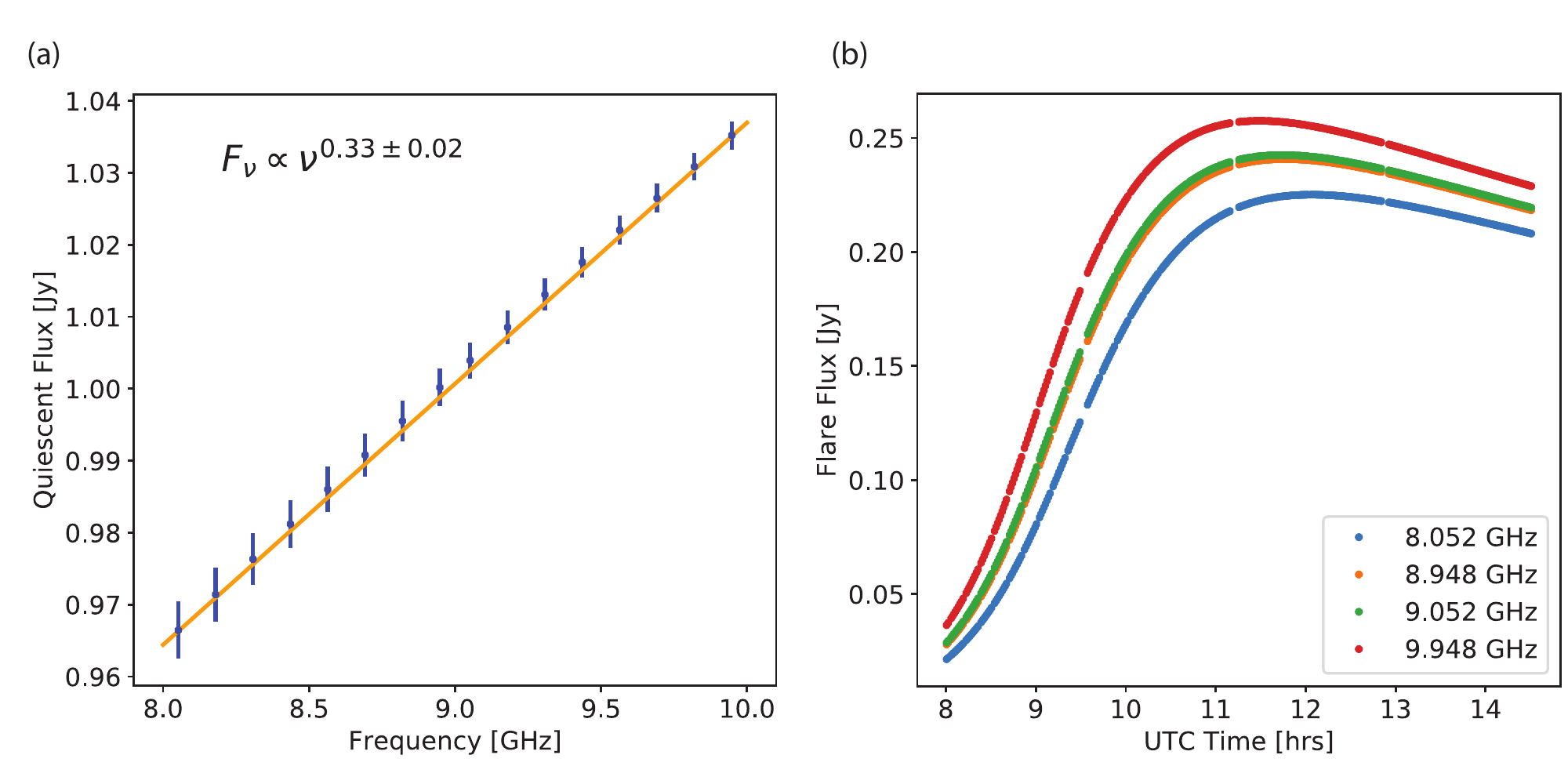}
                \caption{({\color{blue} a}) The fitted quiescent component SED of Sgr A*. The fluxes are calculated at t = 8 hrs UTC, corresponding to the start of the observation. ({\color{blue} b}) The fitted flare light curve for four observed frequencies.}
                \label{fig:adabatic_seds}
            \end{figure*}

        We attempt to simultaneously fit the flaring and quiescent components of Sgr A* assuming the model $S_\nu = S_p(\nu, r) + S_q(\nu, t)$ accurately describes the source. However, we must re-define the value of $r$ as
            \begin{equation}
                r = \left(\dfrac{t - t_s}{\beta t_0}\right)^\beta.
            \end{equation}
        We have introduced a new parameter $t_s$, which is defined as the universal time when the flare begins ($r(t_s) = 0$). This is functionally equivalent to the original definition where $t = 0$ corresponds to the start of the flare. Here, however, $t$ corresponds to the universal time. This offset implies the flare age is equivalent to $t-t_s$. As above, we take $\nu_0 = 9.948$ GHz. 
            \begin{figure}
                \centering
                \includegraphics[trim=1.0cm 1.55cm 2.0cm 1.8cm, clip,width=\columnwidth]{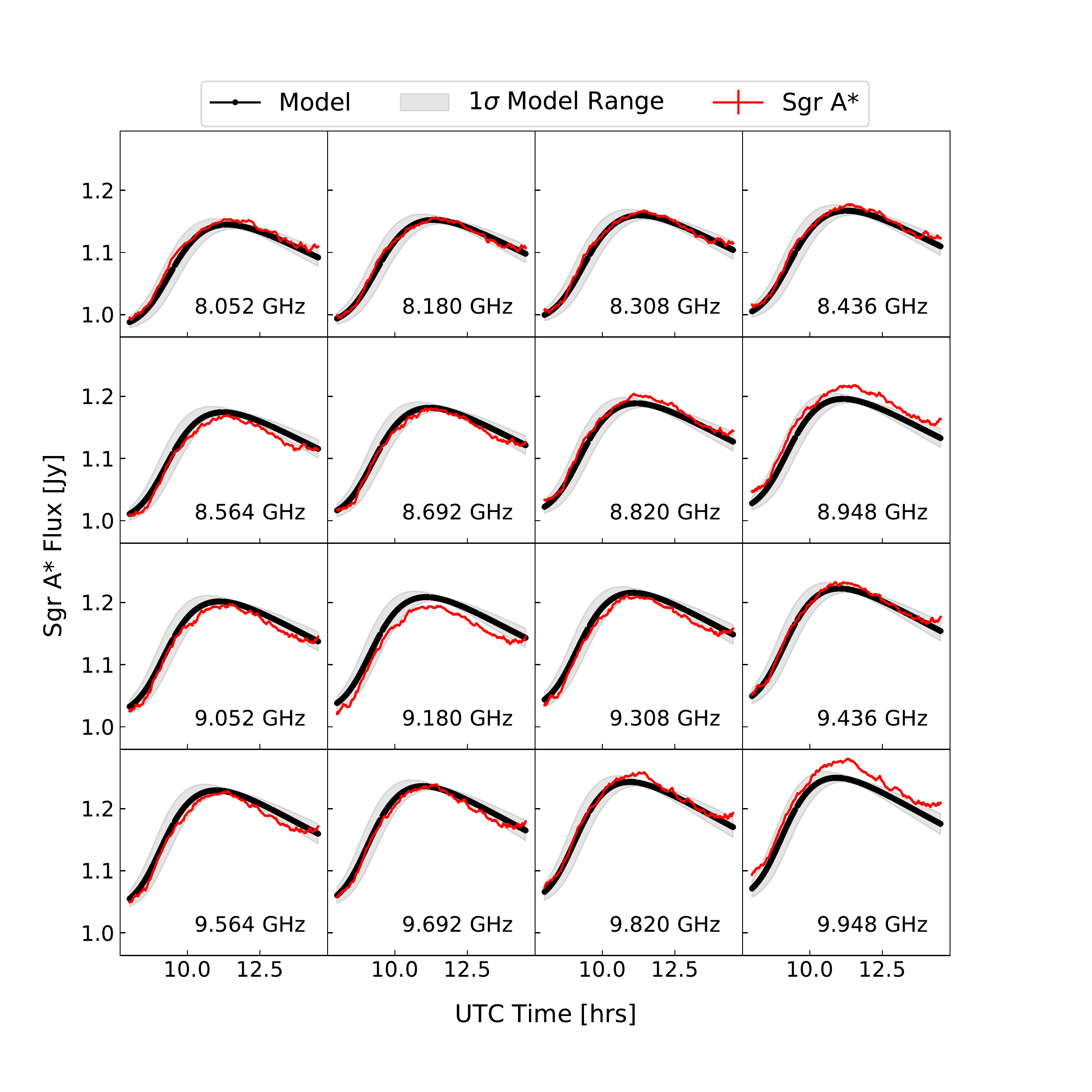}
                \caption{The fitted two-component model for each of the $16$ IFs (in black) compared to the measured Sgr A* light curves (in red). $1\sigma$ errors on the fits are shown in the shaded gray region.}
                \label{fig:Plasmon_fit}
            \end{figure}
        The fitted models are shown in black in Figure \ref{fig:Plasmon_fit} with the $1\sigma$ error range shaded in gray. The original data are plotted for reference in red. We list our fitted parameters with $1\sigma$ errors in Table \ref{tab:Plasmon_params}. In Figure \ref{fig:adabatic_seds}\capcolor{a} we show the fitted quiescent SED, and, in Figure \ref{fig:adabatic_seds}\capcolor{b}, the flare light curves of four frequencies are shown.
        
        At $8.948$ and $9.948$ GHz, the observed flux from Sgr A* is higher than the adiabatic expansion model. This discrepancy is caused by an imperfect flux transfer from the calibrators to Sgr A*, as described in Section \ref{sec:reduction}, and is likely due to an over-correction in the bandpass calibration due to spectral sensitivity roll-off.
            \begin{table}
                \centering
                \begin{tabular}{l|l|r}
                    \hline\hline
                     Parameter & Description & Value \\\hline
                     $S_0^p$ & $9.948$ GHz Peak flare flux  & $0.26 \pm 0.004$ Jy\\
                     $p$  & Flare Energy Power-Index & $0.18 \pm 0.06$\\
                     $\beta$ & Acceleration parameter & $1.50 \pm 0.05$\\
                     $t_0$ & Flare apparent age & $4.24 \pm 0.03$ hrs\\
                     $t_s$ & Start time for blob expansion & $5.13 \pm 0.01$ hrs UTC\\
                     $S_0^q$ & $9.948$ GHz Quiescent flux & $1.04 \pm 0.002$ Jy\\
                     $\dot{S}_q$ & Time derivative of quiescent flux & $-0.0136\pm0.0024$ Jy/hr\\
                     $\alpha_q$ & Quiescent spectral index & $0.33 \pm 0.02$\\\hline
                \end{tabular}
                \caption{Fitted parameters of the two-component model compromised of power-law quiescent and adiabatically-expanding flaring components.}
                \label{tab:Plasmon_params}
            \end{table}

        The value of $\dot{S}_q$ is $-0.0136$ Jy hr$^{-1}$ implying that the quiescent component's flux is decreasing in time. The absolute change in the quiescent flux over the $6$ hour observation is $82$ mJy, which corresponds to about 30\% of the peak flare flux at $9.948$ GHz. The constant offset model (Section \ref{ssec:brute}) cannot fully separate the flaring and quiescent components since flux changes in the latter component are comparable to those in the flaring component.
        
        Previous analyses have assumed constant ($\beta = 1$) expansion of the blob \citep[e.g.,][]{Zadeh2006, Zadeh2008}, where we left it as a free parameter. Our analysis of this flare yields $\beta = 1.50 \pm 0.05$. This is in contrast with the value assumed for Sgr A*; however, it agrees with values found in extragalactic sources \citep{VDL1963}. Future analyses of Sgr A* flaring emission with $\beta$ as a free parameter will be needed to determine its distribution.
               
        For the fitted model, we complete a time delay analysis between the highest and lowest frequencies as in Section \ref{ssec:timedelay}. This yields a lag time $\delta t=20.27^{+2.35}_{-2.66}$ minutes, where the errors quoted are at $2\sigma$. This value is consistent with the observed lags from both techniques presented in Section \ref{ssec:timedelay} to within $2\sigma$. In the adiabatic picture, the time lag between any two frequencies is a function of $t_0$, $t_s$, $\beta$, and $p$. Therefore, this value is an important constraint to determine if the fit was successful. The fitted values are well-constrained due to the number of simultaneous frequency observations that these data provide. With 16 IFs, the model must accurately fit 15 independent time lag combinations, 14 more observations at two frequencies can provide. The 30 second sampling time we use for these light curves gives a large set of data to accurately determine the flare peak flux and, therefore, the quiescent component parameters. In total, our analysis likely provides the most robust set of parameters yet while applying this technique to Sgr A*.
        
       Since the model fit can be extrapolated to all frequencies, we compare the inferred properties of this flare at $22$ and $43$ GHz with those in the literature. We calculate the expected light curves at these two frequencies and determine the estimated time lag is $\delta t_{43-22} = 47.55^{+1.36}_{-0.98}$ minutes. Typical values of $\delta t_{43-22}$ have been observed to be in the range of $20-30$ minutes \citep[e.g.,][]{Zadeh2006, Zadeh2008}. The predicted peak fluxes at $22$ and $43$ GHz are $440$ and $653$ mJy, respectively. \cite{Zadeh2006} reports peak fluxes at these two respective frequencies of approximately $60$ and $137$ mJy, while \cite{Zadeh2008} finds flare peaks between $75-150$ mJy at $22$ GHz and $160-300$ mJy at $43$ GHz. The flare analyzed here is much different from those previously reported, given the difference in predicted peak fluxes and time delays at higher frequencies with narrow bandwidths ($\sim 100$ MHz).
            
       To determine physical parameters, we make assumptions about the electron energy distribution. Previous papers have reported or assumed electron energy indices much steeper than the one found here. For steep power laws, the spectrum normalization can be analytically determined by integrating over all possible electron energies. For $p<1/3$, however, this normalization diverges, and the analytic synchrotron source function gives non-physical values. $p$ found in this analysis is below the analytic limit, and a different normalization must be determined. To obtain physical parameters for this blob, we normalize the spectrum with a maximum electron energy that is $10$ times greater than the characteristic electron energy. This normalization allows us to calculate physical parameters for this source.
       
       We assume the lower and upper electron energy bounds are truncated at $1$ and $100$ MeV, respectively. We adopt equal proton and electron energy densities in the blob, with equal numbers of protons and electrons for charge neutrality. For a peak flux of $260$ mJy at $9.948$ GHz, the blob radius is $1.8\times10^{13}$ cm ($14$ Schwarzschild radii, assuming $M_{\text{Sgr A*}} = 4.3\times10^6$ M$_\odot$), a mass of $5.2\times10^{19}$ g, electron density of $1.3\times10^3$ cm$^{-3}$, and equipartition magnetic field strength of $2.2$ Gauss. With an apparent age of $4.24$ hours, this yields an average expansion velocity of approximately $12000$ km/s, or $0.039$c. The full list of physical parameters is shown in Table \ref{tab:Plasmon_phys}. The absolute values of the physical parameters are accurate to order unity. Most of the uncertainty in these parameters comes from the assumptions regarding the flaring region, such as the upper and lower bounds of the electron energy spectrum. However, our assumptions are very similar to those in the literature. Therefore, a relative comparison between different analyses is still worthwhile.
       
       The derived physical parameters for this flare are much different from those previously reported in the literature. \cite{Zadeh2006} reported a flare with a magnetic field strength of $22$ Gauss and a mass of $4\times10^{19}$ grams with a radius of $4$ R$_\text{S}$ (Schwarzschild radii). \citet{Zadeh2008} fit four Sgr A* flares with the adiabatic expansion model and found magnetic field strengths in the range of $10-76$ Gauss with radii of $0.5-3.2$ R$_\text{S}$. \citet{Zadeh2009} fit four flares \citep[two of which were reported in][]{Marrone2008} and found magnetic field strengths between $13$ and $75$ Gauss with radii of $0.52-9.8$ R$_\text{S}$. The most similar flare to the one reported here had a magnetic field of $13$ Gauss with radius $9.8$ R$_\text{S}$ but with a more steep electron power index of $p=1.5$. \citet{Kunneriath2010} model several simultaneously-observed flares at near-IR and millimeter wavelengths; their results show magnetic fields of $30-86$ Gauss with radii of $0.2-1.3$ R$_\text{S}$. Only one paper cited here calculates the mass of the plasmon, which matches to within an order of unity of our observed value. However, this value is poorly determined as it is calculated using the number density of relativistic electrons. A possible admixture of colder material is not accounted for in the mass estimate.
       
       The flare reported here appears to be an outlier in terms of magnetic field strength, size, and the electron spectrum. The low inferred magnetic field strength and large radius of this plasma compared to earlier work suggests this emission is caused by a transient warm spot in the accretion flow at large radii. \citet{Ressler2020} complete 3D magnetohydrodynamic (MHD) simulations of magnetized Wolf-Rayet stellar winds accreting onto Sgr A*. Their free parameter was plasma $\beta_W$, which they varied to simulate weak and strong magnetic fields in the stellar winds. Their simulations show electron densities $n_e\approx10^{3}$ cm$^{-3}$ at distances of $\approx10-100~R_{S}$ . Their simulations with the lowest $\beta_W$ ($\beta_W = 10^2, 10^4$) yield RMS magnetic field strengths of roughly $1$ Gauss at 250 Schwarzschild radii from the central black hole. These simulations support our hypothesis of a transient warm spot laying far from Sgr A* as the source of this variable emission.
       
       The electron power index is much flatter than those previously reported. This may suggest a relativistic Maxwellian energy distribution for the electrons rather than the power-law distribution assumed here. The Maxwellian distribution is consistent with the transient warm spot hypothesis discussed above, rather than a more energetic event, like a magnetic reconnection. However, attempting to fit the flaring emission with such a model is outside the scope of this paper.
       
       \begin{table}
                \centering
                \begin{tabular}{l|l|r}
                    \hline\hline
                     Parameter & Description & Value \\\hline
                     $E_{min}$ & Electron Lower Energy Bound & $1$ MeV\\
                     $E_{max}$ & Electron Upper Energy Bound & $100$ MeV\\
                     $\tau_{9.948}$ & Optical Depth at $9.948$ GHz & $0.22$\\
                     $n_e$ & Electron density & $1.3\times10^{3}$ cm$^{-3}$\\
                     $R$ & Radius of flaring region & $1.8\times10^{13}$ cm\\
                     $B_{eq}$ & Equipartition magnetic field strength & $2.2$ G\\
                     $M$ & Mass of flaring region & $5.2\times10^{19}$ g\\\hline
                \end{tabular}
                \caption{Physical parameters of the adiabatically-expanding flaring component.}
                \label{tab:Plasmon_phys}
            \end{table}
        
    \subsection{Comparing the Quiescent SED Parameters with Previous Analyses}
        The quiescent spectral index found using the constant offset model is $\alpha = 0.37 \pm 0.04$, while the adiabatic expansion model found $\alpha = 0.33 \pm 0.02$. These values are consistent within $1\sigma$ and are comparable to the $\nu^{0.33}$ power-law spectrum claimed by \citet{Duschl1994}. The quiescent spectral indices here are much steeper than the value \citet{Melia2001} list between $1.36$ and $8.5$ GHz $(\alpha=0.17)$, and is more comparable to their spectral index between $15$ and $43$ GHz ($\alpha=0.3$). Due to the sparse frequency coverage, it is difficult to say whether the change in spectral indices from \citet{Melia2001} is gradual (a bend in the SED) or sudden (a break in the SED). The fluxes at $8.5$ and $15$ GHz in their analysis are $0.8$ and $1$ Jy, respectively, indicating a gradual change. We do not find evidence in our fitted quiescent SED for either of these cases, as it is well-fit by a single power-law. 
        
        We note that the quiescent emission at $9.948$ GHz from this analysis is approximately $10\%$ higher than inferred from the time-averaged SED. This may be suggestive of a different accretion state of the quiescent component compared to the 1980-2000 average presented in \citet{Melia2001}.
        
        
\section{Summary}\label{sec:summary}
    We have reported and analyzed a flare from Sgr A* between $8$ and $10$ GHz from the VLA. Cross correlation analyses find a time delay between the highest ($9.948$ GHz) and lowest ($8.052$ GHz) frequencies on the order of $20$ minutes, which is the first non-zero detection of a time delay within the same radio frequency band. We attempt to determine characteristics of the quiescent and flaring components in two ways. The first subtracts a frequency-dependent constant flux from each light curve, considered to be the quiescent component, and uses the residual flux as a measure of the flaring component light curve. The second uses a sum of a time-dependent power-law, representative of the quiescent component, and adiabatically-expanding synchrotron flux models, representative of the flaring component, to describe Sgr A*. Our fitted adiabatic expansion model reproduces a time delay between the highest and lowest frequency data of approximately $20$ minutes. We have, for the first time, simultaneously characterized the quiescent and flaring components of Sgr A* using the adiabatic expansion model with limited assumptions. The flare analyzed here is abnormal as it is larger in size and has a smaller magnetic field strength than those previously reported in the literature. These physical parameters are consistent with MHD simulations of an accretion flow around Sgr A* that originate from magnetized Wolf-Rayet stellar winds. In this picture, the warm spot in the cold accretion flow is adibatically expanding. The technique presented here will prove useful for future concurrent multi-wavelength observations of Sgr A*, which will allow for characterization of the quiescent and flaring components along a wide swath of frequencies simultaneously.
    
\section*{Acknowledgements}
We dedicate this paper to the recently deceased Roger Hildebrand from the University of Chicago. While never directly mentored by him, JMM is extremely fortunate to have been mentored by many of RH's academic relatives and colleagues throughout his undergraduate and graduate studies. MW and FYZ fondly remember edifying discussions with RH on the polarised emission from dust in the circumnuclear ring and the nature of the magnetic field associated with the Galactic Center filaments. JMM and FYZ thank Marc Royster for an early version of a data set used in this analysis. The authors thank the anonymous referee for their constructive comments. The National Radio Astronomy Observatory is a facility of the National Science Foundation operated under cooperative agreement by Associated Universities, Inc. This research was supported in part through the computational resources and staff contributions provided for the Quest high performance computing facility at Northwestern University which is jointly supported by the Office of the Provost, the Office for Research, and Northwestern University Information Technology.
This analysis has made use of the following Python software packages: \texttt{Jupyter Notebook} \citep{Kluyver2016}, \texttt{pandas} \citep{Mckinney2010}, \texttt{Scipy, Numpy} \citep{SciPy2020}, \texttt{Matplotlib} \citep{Hunter2007} \texttt{uncertainties} \citep{Uncertainties}, \texttt{PYCCF} \citep{Peterson1998, PYCCF}, as well as the following reduction packages outside Python: \texttt{AIPS}, \texttt{CASA} \citep{McMullin2007}, \texttt{ZDCF} \citep{Alexander1997}, \texttt{PLIKE} \citep{Alexander2013}.

\section*{Data Availability}
The data used in this paper are publicly available in their raw form from the NRAO data archive. The reduced data are available upon request.



\bibliographystyle{mnras}
\bibliography{14a232} 

\bsp	
\label{lastpage}
\end{document}